\documentstyle[epsfig, twocolumn]{esapub} 
 
\begin{document} 
 
\def\note #1]{{\bf #1]}}
\def\figdir{./figdir}  
\def\ie{{\rm i.e.}}
\def\muHz{\,\mu{\rm Hz}}
\def\picplace#1{\vspace{#1}}
\def\ddt#1{{\partial #1\over\partial t}}
\def\ddz#1{{\partial #1\over \partial z}}
\def\DDt#1{{{\rm D} #1\over {\rm D} t}}
\def\DD{{\rm D}}
\def\dd{{\rm d}}
\def\vu{\vec{u}}
\def\vxi{\vec{\xi}}
\def\pg{{p_{\rm g}}}
\def\pt{{p_{\rm t}}}
\def\rhobar{{\bar\rho}}
\def\cph{c_{\rm ph}}
\def\ubar{{\bar u}}
\def\uzprime{u_z^\prime}
\def\pbar{{\bar p}}
\def\pgbar{{\bar p}_{\rm g}}
\def\uzbar{{\bar u_z}}
\def\bra{\langle}
\def\ket{\rangle}
\def\Sect#1{Section \ref{#1}}
\def\Eq#1{Eq.\ (\ref{#1})}
\def\Fig#1{Fig.\ \ref{#1}}
\def\apj{{\rm ApJ}}
\def\s{\, {\rm s}}
\def\km{\, {\rm km}} 
 
\setlength{\parindent}{0pt} 
\setlength{\parskip}{ 10pt plus 1pt minus 1pt} 
\setlength{\hoffset}{-1.5truecm} 
\setlength{\textwidth}{ 17.1truecm } 
\setlength{\columnsep}{1truecm } 
\setlength{\columnseprule}{0pt} 
\setlength{\headheight}{12pt} 
\setlength{\headsep}{20pt} 
\pagestyle{esapubheadings} 
 
\title{\bf TESTS OF CONVECTIVE FREQUENCY EFFECTS WITH SOI/MDI HIGH-DEGREE DATA} 
\author{{\bf C.S. Rosenthal$^1$, J.\ Christensen-Dalsgaard$^{2,3}$,
A.G.\ Kosovichev$^{4}$, {\AA}.\ Nordlund$^{2,5}$,}\\{\bf J.\ Reiter$^{6}$,
E.J. Rhodes\ Jr$^{7,8}$, J.\ Schou$^{4}$, R.F.\ Stein$^{9}$, R.\ Trampedach$^{3}$}
\vspace{2mm} \\ 
$^1$High Altitude Observatory/NCAR, P.O. Box 3000, Boulder, CO 80307-3000 \\ 
$^2$Teoretisk Astrofysik Center, Danmarks Grundforskningsfond\\ 
$^3$Institut for Fysik og Astronomi, Aarhus Universitet, DK 8000 Aarhus C, Denmark\\ 
$^4$W.W. Hansen Experimental Physics Laboratory, Stanford University, Stanford, CA 94305-4085,
USA\\ 
$^5$Niels Bohr Institutet for Astronomi, Fysik og Geofysik, Astronomisk Observatorium, 
K\o benhavns\\ Universitet, Juliane Maries Vej 30, DK 2100  K\o benhavn \O, Denmark\\ 
$^6$Mathematisches Institut, Technische Universit\"at M\"unchen, D-80333 M\"unchen, Germany\\ 
$^7$Department of Physics and Astronomy,University of Southern California, Los Angeles, CA
90089, USA \\ 
$^8$Space Physics and Astrophysics Research Element, Jet Propulsion Laboratory, \\
California Institute of Technology, 4800 Oak Grove Drive, Pasadena, Ca 91109-8099, USA\\
$^9$Department of Physics and Astronomy, Michigan State University, East Lansing, MI 48824,
USA\\ 
} 
 
\maketitle 
 
\begin{abstract} 
 
Advances in hydrodynamical simulations 
have provided new insights into the effects of
convection on the frequencies of solar oscillations.
As more accurate observations become available,
this may lead to an improved understanding of the
dynamics of convection and the interaction between
convection and pulsation (\cite{Rosenthal+99}). 
Recent high-resolution observations from the
SOI/MDI instrument on the SOHO spacecraft have 
provided the so-far most-detailed observations of 
high-degree modes of solar oscillations, which are
particularly sensitive to the near-surface properties of the Sun.
Here we present preliminary results of a comparison
between these observations and frequencies computed
for models based on realistic simulations of near-surface convection.
Such comparisons may be expected to help in identifying the
causes of the remaining differences between the observed
frequencies and those of solar models. \vspace {5pt} \\ 
 

  Key~words: solar oscillations; convection.

\end{abstract} 
 
\section{INTRODUCTION} 
 
In stellar-structure theory,
the standard Mixing-Length Model of convection (MLT)
continues to be the industry standard, despite its many limitations. For example
MLT cannot account for the broadening of photospheric spectral
lines, presumably due to overshooting, nor can it model the effect of 
turbulent pressure on the mean structure in a consistent way. More subtly
perhaps, the reduction of the true 3-D radiation hydrodynamics to 1-D
hydrostatic stratification
eliminates some significant contributing effects including the role of
correlations, which result in a mean structure which does not
satisfy the locally-valid equation of state, and radiative effects due 
to opacity inhomogeneities, which 
result in hotter surface layers for a given effective temperature (see
\cite{Stein+Nordlund98b}).

Helioseismology has considerable potential as a tool for testing the accuracy
of numerical models of these surface regions through comparison of
predicted and measured frequencies. A major contribution is the effect of
turbulent pressure on the mode frequencies (\cite{Kosovichev95}).
In such a comparison, it is vital to distinguish between {\bf
model} effects on frequencies, resulting
from errors or inaccuracies in the mean structure, and {\bf modal} effects resulting
from uncertainties in the physics of wave propagation in these turbulent layers. 
We use
numerical simulations of the surface layers to handle model 
effects while studying various physically-motivated assumptions about the modal
physics. 

\section{DATA ANALYSIS} 

Our data analysis procedure is described in more detail by 
\nocite{Rhodes+97,Rhodes+98} Rhodes et al. (1997, 1998).
A 60.75 day time series of MDI Full-Disk
Dopplergrams, beginning May 23, 1996, was used. After limited gap-filling, the final duty cycle was 97.3~\%. The so-called 
``averaged-spectrum'' method is used to obtain frequency estimates, with results from
low- and intermediate-degree power spectral peaks being used to obtain a correction due
to asymmetry which is then applied to the frequencies obtained by ridge-fitting at 
higher degrees. Results are obtained for degrees up to $\ell=1000$.

\section{THE STANDARD SOLAR MODEL}

The standard solar model we use is Model~S of \cite*{JCD+96science}. It uses 
the OPAL equation of state (\cite{Rogers+96}), OPAL opacities (\cite{Iglesias+92}) 
at high temperature and 
(\cite{Kurucz91}) opacities
in the atmosphere.
The atmosphere is determined by a 
$T-\tau$ relation (where $T$ is temperature and $\tau$ is optical depth)
fitted to the HSRA reference atmosphere (\cite{HSRA}) and standard 
MLT (\cite{Boehm-Vitense58}) is used for the 
convection zone. The
effects of turbulent pressure are not included in Model~S.

Figure 1 shows a comparison between adiabatic eigenfrequencies of Model~S and the
data. The frequency residuals have been scaled with the quantities $Q_{nl}$ which
represent the ratio of the mode inertia of a particular mode to that of a radial
mode of the same frequency. Mode inertias were normalised at the photosphere, defined
as the location where the temperature is equal to the effective temperature.
It is evident that, except for modes of low order, the
scaled residuals are essentially a function of frequency alone, an indicator that
the predominant effect responsible for the residuals lies close to the solar
surface (\cite{JCD+Berthomieu91}). 
 
 \begin{figure}[h] 
  \begin{center} 
    \leavevmode 
  \centerline{\epsfig{file=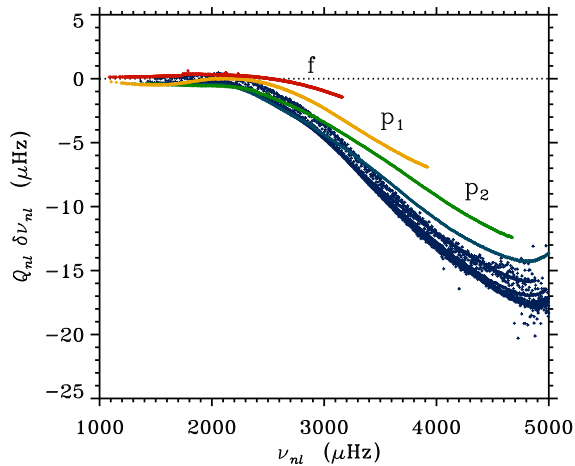,width=8.0cm}} 
  \end{center} 
  \caption{\em  Measured frequency residuals in the sense
(observations) --
(model), scaled by $Q_{nl}$ for selected modes in the range $0\le\ell\le 1000$.
The computed frequencies are for Model~S of Christensen-Dalsgaard et al. (1996).
Radial orders of the lowest-order modes are indicated $f$, $p_1$, and $p_2$.} 
  \label{fig:residuals_S} 
\end{figure} 

\section{MODEL CONTRIBUTIONS AND NUMERICAL SIMULATION}

We treat the model contributions to the residuals by using radiation-hydrodynamic 
simulations of the surface layers to construct new mean models,
by patching temporal and spatial averages of the simulations to convective 
envelope models constructed with standard physics. The simulations are described
in detail by \nocite{Stein+Nordlund89b,Stein+Nordlund97a} Stein \& Nordlund (1989,
1998). 

Our philosophy is to construct numerical simulations and standard MLT envelope
models each using the same physics, thus allowing us to isolate the effects
of 3-D convective dynamics.  
In this case we used simulations calculated with the MHD equation
of state (\cite{MHD88a,MHD88b,MHD88c,MHD90}) and Kurucz opacities. The
effective temperature of the simulation was 5777K.

A standard envelope model was constructed with the same physics, and 
using standard MLT with the addition of a 
parameterised turbulent pressure of the form $\pt=\beta\rho v^2$,
where $\rho$ is density and $v$ is the 
the convective velocity, obtained from the mixing-length formulation.
The extra parameter $\beta$ and the usual mixing length parameter 
were then adjusted
to allow a continuous matching of density, pressure and temperature. Because the
matching point is at a depth where the turbulent pressure is already very small (of
order 1\% of the total pressure) we do not consider this matching process to be a
useful parameterisation of the relation between mixing-length velocity and turbulent
pressure in general.
Strikingly, the resulting matched model has a convection zone 
depth of $0.286R_\odot$, very close to
the helioseismic value of $(0.287 \pm 0.003) R_\odot$ (\cite{JCD+91,Kosovichev+Fedorova91}) 
or $(0.287 \pm 0.001) R_\odot$
(\cite{basu:dCZ}).
This is a remarkable, parameter-free validation of the simulations.

\section{MODAL CONTRIBUTIONS}

The process described above yields a hydrostatic pressure and density stratification
for a patched model based on the simulation.
We consider only adiabatic oscillations of this patched model.
Under the assumption of adiabaticity, the modal effects are essentially
confined to the relation between the Lagrangian perturbations $\delta p$
and $\delta \rho$ in pressure and density,
\begin{equation}
{\delta p\over p}= \Gamma_1 {\delta\rho\over\rho} \; .
\end{equation}
When convection is ignored,
$\Gamma_1 = (\partial \ln p /\partial \ln \rho)_s$,
the derivative being at constant specific entropy $s$,
is simply the thermodynamic adiabatic exponent.
Here we incorporate modal effects by the use
of different recipes for $\Gamma_1$; 
in particular, they define the adiabatic sound speed 
$c = (\Gamma_1 p /\rho)^{1/2}$.
We consider two possibilities:
the Gas Gamma Model (GGM) and the Reduced Gamma Model (RGM).

In the GGM we simply use the horizontal and temporal mean value of 
the thermodynamic $\Gamma_1$ from the
simulations. In the RGM, this is replaced by the so-called ``reduced gamma'':
\begin{equation}
\Gamma_1^{(r)}\equiv {\left<\Gamma_1p_{\rm g}\right>\over p} \; ,
\label{eq:red_gamma}
\end{equation}
where $p_{\rm g}$ is the gas pressure and $p$ is the total pressure. The reduced gamma 
was originally introduced
by \cite*{Rosenthal+95b}, who argued
that if the Lagrangian perturbation to the turbulent pressure,
$\delta \bar \pt$, were zero then
\begin{equation}
{{\delta \bar p}\over p}={ {\delta\bar \pg} \over \bar p} =
\Gamma_1^r {\delta\bar\rho\over\bar\rho} \; .
\end{equation}
\cite*{Rosenthal+99} show that the definition (\ref{eq:red_gamma}) is the correct
one to use if $\delta \bar \pt=0$ provided certain other conditions are also met, 
principally that the
time variation of the convective energy fluxes can be neglected.

We also construct a Standard Envelope Model (SEM) which uses the same physics as
the simulations, has no turbulent pressure, and is calibrated to have the same
convection-zone depth as the patched models. The SEM is constructed using the Rosseland
Mean of the opacities used in the simulations and a $T-\tau$ relation determined from the
averaged simulation itself. Thus the differences between the models should isolate
truly convective effects.

\section{MODEL DIFFERENCES}

\cite*{JCD+Thompson97} have shown that when model
differences are expressed in Lagrangian terms (i.e., on a fixed mass
scale) the frequency change due to near-surface perturbations can be written
approximately as 
\begin{equation}
{\delta \omega_{nl} \over \omega_{nl}} \simeq
\int_0^R \tilde K_{\upsilon,c}^{nl}(r) {\delta_m \upsilon \over \upsilon}
{\rm d} r \; ,
\end{equation}
where ${\delta_m \upsilon / \upsilon}$ is the Lagrangian perturbation to
the quantity $\upsilon=\Gamma_1/c$, and where the kernels $\tilde
K_{\upsilon,c}^{nl}(r)$ can be calculated
from the structure and eigenfunctions of the reference model.
For an isothermal layer, $\upsilon=2\omega_c/g$, where $\omega_c$ is the acoustic
cutoff frequency. Hence changes in $\upsilon$
correspond physically to changes in the size of the acoustic cavity.

Figure 2 shows Lagrangian differences in $\upsilon$ between the various models. The
GGM-SEM differences largely reflect changes in density due to the elevation of the
photospheric regions by turbulent pressure. In addition, the RGM-SEM differences show
the effect of the turbulent reduction of $\Gamma_1$. Comparing (GGM-SEM) with (GGM-Model~S) 
we can see
that the different 
physics between SEM and Model~S (principally OPAL v. MHD
equation of state and a different $T-\tau$ relation) produces further small changes 
confined extremely close to the surface.

 \begin{figure}[h] 
  \begin{center} 
    \leavevmode 
  \centerline{\epsfig{file=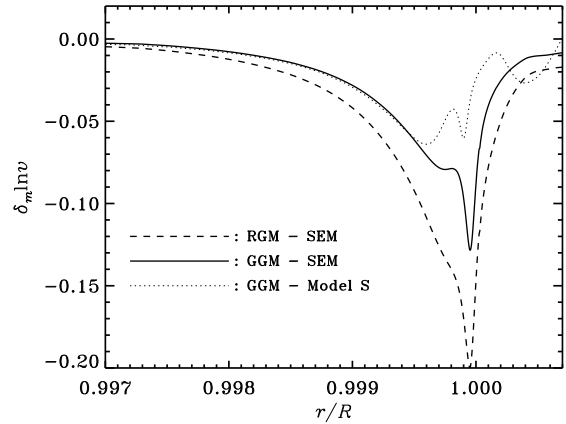,width=8.0cm}} 
  \end{center} 
  \caption{\em  Logarithmic Lagrangian 
(at fixed mass)
differences in $\upsilon = \Gamma_1/c$
between patched and comparison envelope models (solid and
dashed lines): SEM is the comparison model, using mixing-length theory,
while GGM and RGM are patched models assuming the
averaged thermodynamic $\Gamma_1$ and the reduced $\Gamma_1^r$,
respectively.
Also shown is the difference between the GGM envelope and standard solar model
(Model~S).} 
  \label{fig:model_diffs} 
\end{figure} 

\section{OSCILLATION FREQUENCIES}

We restrict ourselves to modes trapped within the convection zone. This allows
us to include direct comparisons also with Model~S. Figure 3 shows that the frequency differences
between GGM and SEM are similar
in magnitude and shape to those between the data and Model~S (Figure 1), whereas
corresponding differences between RGM and SEM are much larger than required.
GGM-Model~S differences are broadly similar to GGM-SEM differences.

 \begin{figure}[h] 
  \begin{center} 
    \leavevmode 
  \centerline{\epsfig{file=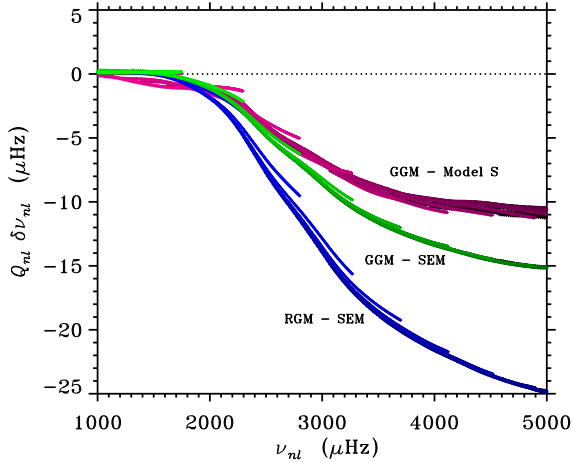,width=8.0cm}} 
  \end{center} 
  \caption{\em  Scaled differences
between frequencies of patched (RGM and GGM) and comparison (SEM) envelope models 
and between the GGM envelope and the standard solar model (Model~S).} 
  \label{fig:freq_diffs_1} 
\end{figure} 

 \begin{figure}[!ht] 
  \begin{center} 
    \leavevmode 
  \centerline{\epsfig{file=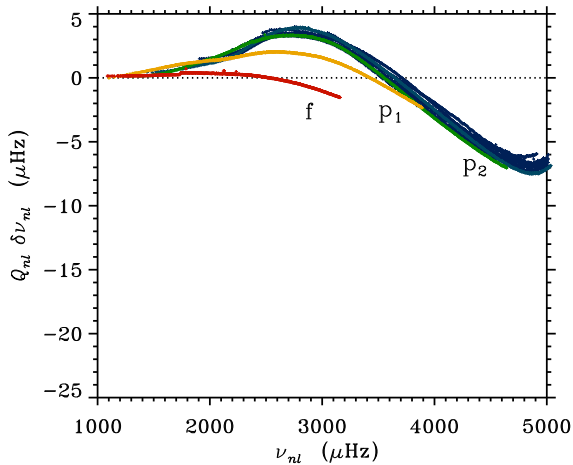,width=8.0cm}} 
  \end{center} 
  \caption{\em Measured frequency residuals in the sense
(observations) -- (model),
scaled by $Q_{nl}$. The model frequencies are for
the GGM, with the (unreduced) gas $\Gamma_1$.} 
  \label{fig:freq_diffs_2} 
\end{figure} 

Figure 4 is a direct comparison between GGM frequencies and the data. The residuals
are much reduced compared to Figure 1. However the ``bump'' below 3mHz suggests that
the combined model and modal differences in Figure 2 are too deeply penetrating. Thus
the Gas Gamma hypothesis cannot be a completely accurate description of what is
occurring.

\section{INTERPRETATION AND CONCLUSIONS}

Our statement that the remaining discrepancy in frequency must be due to modal,
rather than model, effects is based on the belief that the simulations are
fundamentally correct. Why do we have such faith in the simulations? The principal 
model effect on the frequencies is the elevation of the photospheric layers
by the combined effect of turbulent pressure and an effect of the 3-D
dynamics, whereby hotter regions, which have higher opacity, contribute less
to the emitted radiation. The higher 
mean temperature of the 3-D model is thus ``hidden'' from view, but is 
reflected in the pressure stratification.

We have carried out a number of convergence tests on the simulations involving
runs at various resolutions. The thermal stratification is rather insensitive
to numerical resolution ({\em cf}\/. \cite{Stein+Nordlund97a}, Fig.\ 26). The
variation of turbulent pressure with resolution is shown in Figure 5.

 \begin{figure}[h] 
  \begin{center} 
    \leavevmode 
  \centerline{\epsfig{file=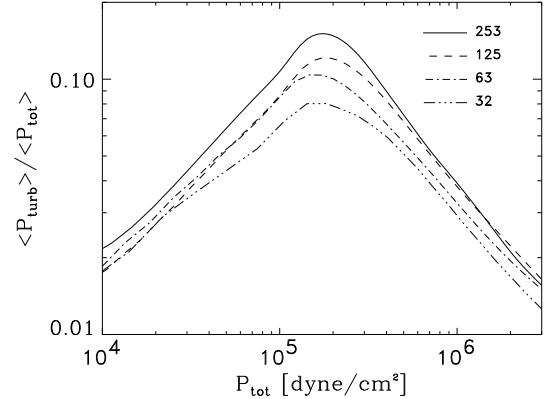,width=8.0cm}} 
  \end{center} 
  \caption{\em Mean turbulent pressure 
divided by total (gas + turbulent) pressure as a
function of depth for four different runs, with different numerical
resolutions: $253^2 \times 163$, $125^2 \times 82$, $63^2 \times 63$, 
and $32^2 \times 41$. (Note that the total duration of the $253^2 
\times 163$ run is only 10 minutes, while we estimate that a minimum 
duration of 30--60 minutes is required for statistically stable 
results.)} 
\end{figure} 

The highest-resolution runs already produce the correct turbulent broadening,
asymmetries, and blue-shift of spectral lines and so the maximum turbulent 
pressure cannot increase much beyond the largest values found so far
(\cite{Nordlund+Dravins90,Dravins+Nordlund90a,Dravins+Nordlund90b}). In fact,
despite the apparent steady increase of turbulent pressure with numerical
resolution shown in Figure 5, the density structure varies very little 
between the various runs. Hence, we expect that the elevation of the 
atmosphere is nearly fully accounted for at our highest resolutions.

What then are the nature of the modal effects which account for 
the remaining frequency differences? One clue is provided by
\cite*{Stein+Nordlund91b}, who show that the time variation of
the turbulent pressure is related to the compression in a much more
complicated fashion than that implied by either the GGM or the RGM. 
In practice the turbulent component of gamma,
$\partial\ln \pt/\partial\ln \rho$, can
be negative, and is a strong function of height. Evidently, we need to use 
such information from the simulations themselves to determine a better
model for $\Gamma_1$ than the two simple cases considered here. 

However, the limitations even of this approach are evident from consideration
of the f modes. These are almost entirely insensitive to the mean solar structure and
so no treatment based on variations in either $\Gamma_1$ or the hydrostatic structure
will ever improve the agreement between measurement and theory in their case. The
principal contribution to the f-mode residuals probably comes from the advection
of wavefronts by turbulent motions 
(\cite{Murawski+Roberts93a,Murawski+Roberts93b,Gruzinov98}),
an effect which has proved extremely difficult to model realistically. 
However, \cite*{Murawski+98} have derived a dispersion relation
for the f mode in a simple model containing a random velocity field,
and shown that the frequency shift and linewidth of the f mode
for this model are consistent with the properties obtained from the
high-resolution MDI data. It is likely that turbulent advection results in
frequency shift of the p modes in addition to the turbulent pressure
effect.
It may be, once
again, that we will have to turn to the simulations to provide us with clues
as to how to describe these processes.

\section*{ACKNOWLEDGMENTS}

The National Center for Atmospheric Research is sponsored 
by the National Science Foundation. CSR acknowledges support from 
SOI/MDI NASA GRANT NAG5-3077.
JC-D, \AA N and RT acknowledge support of Danmarks Grundforskningsfond
through the establishment of the Theoretical Astrophysics Center.
RFS acknowledges support from NASA grant NAG5-4031 and NSF grant 9521785.

\bibliographystyle{aabib}
\bibliography{aajour,convection,oscillations,Aake,books,jcd,Colin}

\end{document}